\documentclass[fleqn,usenatbib]{mnras}
\usepackage{amssymb}
\usepackage{graphicx}
\usepackage{amsmath}
\usepackage{xspace}
\usepackage[T1]{fontenc}
\usepackage{ae,aecompl}
\usepackage{newtxtext,newtxmath}
\usepackage{natbib}
\usepackage{anyfontsize}

\def\Msun{\hbox{$\rm\thinspace M_{\odot}$}}
\newcommand{\Teff}{\mbox{$T_{\rm eff}$}\xspace}
\newcommand{\Lbol}{\mbox{$L_{\rm bol}$}\xspace}
\newcommand{\Lsun}{\mbox{$L_{\odot}$}\xspace}
\newcommand{\Mjup}{\mbox{$M_{\rm Jup}$}\xspace}
\newcommand{\orcid}[1]{\href{https://orcid.org/#1}{\textsuperscript{\includegraphics[width=8pt]{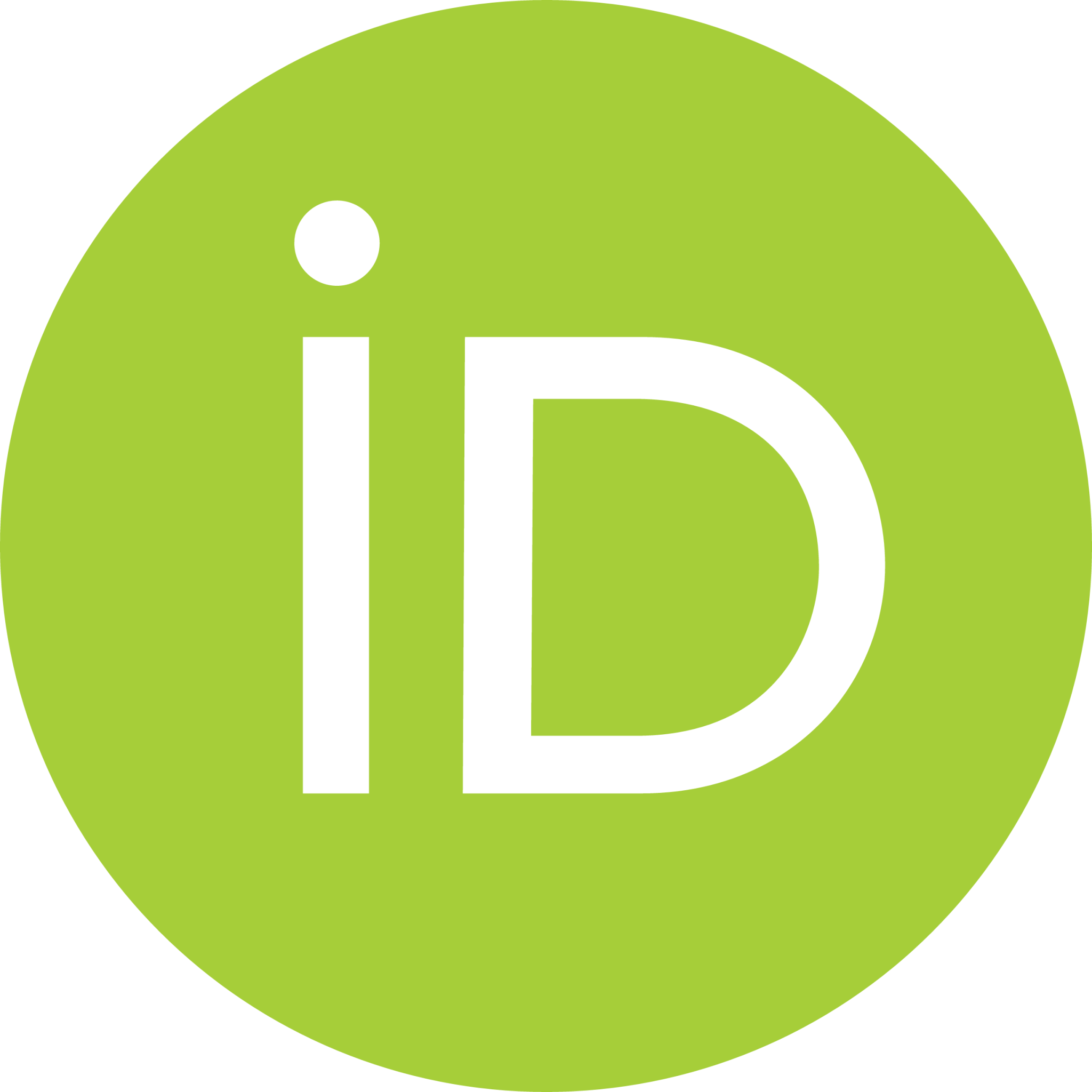}}}}

\title[Proper motion confirms that Capotauro is a Y-dwarf]
{A clear detection of proper motion confirms that the claimed $\mathbf{z\simeq32}$ galaxy candidate, ``Capotauro'', is a Y-type brown dwarf}
\author[F.-Y. F. Liu et al.]
{F.-Y. F. Liu$^{1}$\orcid{0000-0003-1386-3676},$^{1}$\thanks{E-mail: Fengyuan.Liu@ed.ac.uk}, D. J. McLeod$^{1}$\orcid{0000-0003-4368-3326}, B. J. Sutlieff$^{1,2}$\orcid{0000-0002-9962-132X}, T. J. Dupuy$^{1,2}$\orcid{0000-0001-9823-1445}, J. S. Dunlop$^{1}$\orcid{0000-0002-1404-5950}, X. Chen$^{1}$\orcid{0009-0005-9339-2369}, \newauthor R. J. McLure$^{1}$\orcid{0009-0005-9742-2318}, H.-H. Leung$^{1}$\orcid{0000-0003-0486-5178}, S. Antonogiannaki$^{1}$\orcid{0009-0006-2644-6424}, C. Bondestam$^{1}$\orcid{0009-0008-3775-5112
}, E. Laplace$^{3,4}$\orcid{0000-0003-1009-5691}, R. Begley$^{5}$\orcid{0000-0003-0629-8074}, \newauthor A. C. Carnall$^{1}$\orcid{0000-0002-1482-5818}, F. Cullen$^{1}$\orcid{0000-0002-3736-476X}, C. T. Donnan$^{6}$\orcid{0000-0002-7622-0208}, R. S. Ellis$^{7}$\orcid{0000-0001-7782-7071}
\\
$^{1}$Institute for Astronomy, University of Edinburgh, Royal Observatory, Blackford Hill, Edinburgh, EH9 3HJ, UK\\
$^{2}$Centre for Exoplanet Science, University of Edinburgh, Edinburgh, EH9 3HJ, UK \\
$^{3}$Institute of Astronomy, KU Leuven, Celestijnenlaan 200D, B-3001 Leuven, Belgium\\
$^{4}$Leuven Gravity Institute, KU Leuven, Celestijnenlaan 200D, box 2415, 3001 Leuven, Belgium\\
$^{5}$ Armagh Observatory and Planetarium, College Hill, Armagh, BT61 9DG, N. Ireland, UK\\
$^{6}$ NSF's National Optical-Infrared Astronomy Research Laboratory, 950 N. Cherry Avenue, Tucson, AZ 85719, USA \\
$^{7}$ Department of Physics and Astronomy, University College London, Gower Street, London WC1E 6BT, UK\\}

\date{Accepted XXX. Received YYY; in original form ZZZ}

\pubyear{\the\year{}}

\begin{document}
\label{firstpage}
\pagerange{\pageref{firstpage}--\pageref{lastpage}}
\maketitle

\begin{abstract}
The compact red source ``Capotauro'', discovered in deep {\it JWST} imaging in 2025, has been suggested as a possible galaxy candidate at redshift $z\simeq32$ on account of its extreme NIRCam colours. However, given the lack of evidence that this source is spatially resolved, a very cool brown dwarf, at a distance consistent with the scale-height of the Milky Way stellar disc ($\simeq 500$\,pc), provides an alternative explanation of the observed photometry. Here we exploit new medium-band NIRCam imaging obtained $\simeq 3.5$ years after the discovery data to test this hypothesis. Using filter-dependent point-spread-function fitting, a robust local relative astrometric frame defined by 68 compact reference sources, and a joint four-image injection--recovery analysis, we show that Capotauro has moved by $132 \pm 20$\,mas over the 3.5-yr interval span of the observations, ruling out the possibility that it is an extragalactic source (galaxy, AGN, or supernova) at $> 6 \sigma$. The observed apparent proper motion is $(\mu_\xi,\mu_\eta)=(+9.4^{+6.4}_{-4.8},-35.6^{+5.1}_{-5.9})$\,mas\,yr$^{-1}$, or $37.6^{+5.5}_{-5.6}$\,mas\,yr$^{-1}$ in total. Through comparison of its photometric spectral energy distribution with empirical templates, we find that Capotauro is best described as a brown dwarf of spectral type Y$1.0\pm0.5$ ($\Teff\approx350$\,K) at a distance of $730\pm110$\,pc (although the existing data and spectral templates are insufficient to rule out an even colder, later-type brown dwarf). Capotauro is thus one of the most distant Y~dwarfs found to date, as expected given its discovery in a deep {\it JWST} extragalactic survey field. This result demonstrates the value of multi-epoch imaging for identifying substellar contaminants among the most extreme photometric-redshift candidates. 
\end{abstract}
\begin{keywords}
brown dwarfs -- proper motions -- astrometry -- galaxies: high-redshift -- galaxies: evolution -- galaxies: formation
\end{keywords}

\section{INTRODUCTION}

The first four cycles of \textit{JWST} operations have revolutionised our understanding of early galaxy evolution, with the selection of large samples of robust $z=10$--11 galaxy candidates with NIRCam now becoming routine (see, e.g., \citealt{finkelstein2024,Donnan2023a,mcleod2024}), and the spectroscopic redshift frontier having advanced to $z \simeq 14$ \citep{carniani2024,naidu2026}. 

This dramatic observational progress has enabled the statistical study of galaxy evolution to be extended back to within $\simeq 300$\,Myr of the Big Bang via the determination of the evolving galaxy UV luminosity function (UV LF) out to $z \simeq 14$ from the ever-growing database of deep+wide NIRCam imaging surveys (e.g., \citealt{Franco2025,Weibel2026,mcleod2026}).
While a consensus has been reached regarding the smooth and relatively slow decline of the UV LF (and hence star-formation rate density $\rho_{\rm SFR}$) out to at least $z\simeq12$ \citep{Harikane2023,mcleod2024,finkelstein2024,donnan2024}, there is now growing evidence for a more rapid descent in star-formation activity towards yet earlier times (i.e., at $z >12$:  \citealt{Weibel2026,mcleod2026}). However, some studies have reported a surprisingly high abundance of $z>14$ candidates, sometimes from relatively modest search areas \citep{castellano2025,hainline2026a, gandolfi2026b}, and in extreme cases potentially pushing the redshift frontier out to $z>20$ \citep{perezgonzalez2025}. Such claims have strong implications for our understanding of early galaxy evolution (and indeed cosmic hydrogen reionisation), given that some of the reported galaxy number densities at $z>20$ would imply extensive (and incredibly efficient) star formation $< 200$\,Myr after the Big Bang. 

This new debate over the very early evolution of the UV LF (and hence $\rho_{\rm SFR}$) might appear reminiscent of the now-settled, pre-\textit{JWST} dispute over the evolution of $\rho_{\rm SFR}$ at $z>8$ (see, e.g., \citealt{McLeod2015,McLeod2016,Oesch2018}). However, the current situation is in fact very different, because the red wavelength coverage offered by {\it JWST} in principle enables the discovery of Lyman-break galaxy candidates out to extremely high redshifts (given sufficient integration time). A typical NIRCam extragalactic survey usually provides imaging in at least \textit{four} broad-band filters (F200W, F277W, F356W, F444W) of the putative Lyman break at $z=15$ (three for $z=20$) and it is not uncommon for at least one medium band, such as F410M, to be included, particularly in legacy surveys such as PRIMER \citep{dunlop2021}, JADES \citep{rieke2023, eisenstein2023} and PEARLS \citep{Windhorst2023}. Thus, it is perhaps no surprise that some recent studies have begun to push Lyman-break selection out towards F356W-dropouts, corresponding to $z>30$. Such searches are aided by the fact that secure detection of an object in both F410M and F444W can help mitigate the risk that single-band detections (longward of the putative Lyman break) could be contaminated by artefacts and cosmic rays. 

The recent discovery of the F356W-dropout candidate ``Capotauro'' \citep[CEERS ID U-100588,][]{gandolfi2026} (ICRS J2000: RA 14:19:32.97, Dec +52:47:52.11) in the \textit{JWST} CEERS survey \citep{Finkelstein2022,Finkelstein2023,Bagley2023} was thus met with much interest/excitement. The strength ($\simeq 3$\,mag.) of the apparent spectral break in this object yields an apparently secure and unique photometric redshift solution of $z_{\rm phot}\simeq32$, robust against lower-redshift galaxy interlopers unless one fits a solution with an extreme (and arguably unphysical) combination of dust reddening and strong emission lines \citep{gandolfi2026}. Moreover, the NIRSpec spectrum of Capotauro, observed as part of the CAPERS programme (GO~6368; PI~Dickinson; see \citealt{Donnan2025b}), corroborated the lack of detection at $<3\,\mu$m, albeit with a potential detection of flux around $\simeq3.5\,\mu$m.

The competing interpretation, also explored by \citet{gandolfi2026}, is that Capotauro could be a cool ($\leq300$\,K) substellar Y-type brown dwarf at a distance of several hundred parsec. Due to their intrinsic faintness, the vast majority of known Y~dwarfs are at distances of $<30$\,pc \citep[e.g.,][]{Kirkpatrick2021, Fontanive2025}. The Y-dwarf interpretation for Capotauro has been favoured by the more recent study by \citet{hainline2026b}, who derive excellent fits to Capotauro's photometry consistent with a $\sim$350~K object at a distance of $\sim$0.5~kpc. They show that this source appears to be similar to other objects in their sample of cool brown dwarfs, and highlight the need for caution that such brown dwarfs could be contaminating extreme redshift ($z>30$) samples. Indeed, several recent searches have uncovered brown dwarfs in extragalactic survey data obtained with {\it JWST}, out to distances of a few kpc \citep[][]{Langeroodi2023, Burgasser2024, Hainline2024a, Hainline2024b, Li2026, Morrissey2026}. Although relatively few Y~dwarfs have been discovered to date, such objects are of critical importance for improving our understanding of the (sub)stellar initial mass function \citep[e.g.,][]{Kirkpatrick2012, Burningham2013, Hainline2024a, hainline2026b}. Furthermore, Y~dwarfs have typical masses \citep[$\sim$3--25\,\Mjup, e.g.,][]{Luhman2011, Dupuy2013, Leggett2026} and temperatures \citep[$\lesssim$500\,K, e.g.,][]{Leggett2021, Fontanive2026} comparable to those of the coldest directly imaged giant exoplanets, making them valuable proxies for studying these objects \citep[e.g.,][]{Kuzuhara2013, Matthews2024, BardalezGagliuffi2025, Lagrange2025, Crotts2025, Gibbs2026, Sutlieff2026}.

Capotauro is one of several objects which have presented an intriguing substellar versus extreme-redshift-galaxy dichotomy. For instance, in the region around the Bullet cluster, \citet{bradac2026} identified two potentially ultra-high-redshift candidates with $z_{\rm phot}\geq30$ in NIRCam imaging, but then found, using subsequent NIRSpec observations, that the spectra of both objects are also consistent with cold Y-dwarfs (272--351\,K and 445--525\,K). Further NIRCam imaging taken approximately one year later then provided secure proper motion detections, confirming that both objects are indeed Y-dwarfs \citep{bradac2026}. Notably, the NIRSpec spectrum and estimated temperature of the colder of these two objects, F356W-dropout Bullet-BD1, is very similar to that of Capotauro, in the scenario that it is a Y~dwarf \citep{gandolfi2026}. Understanding this dichotomy and how we can best distinguish between these two scenarios, as well as characterise these objects, is therefore of critical importance for future studies of high-redshift galaxies {\it and} brown dwarfs, as well as for planning future deep observations with \textit{JWST}.

In this paper, we use new (i.e., second-epoch) NIRCam imaging to reveal the unambiguous detection of proper motion for Capotauro and also use the expanded (broad-band + medium-band) NIRCam dataset to re-analyse its spectral energy distribution. Our results confirm that it is a Y-type brown dwarf within the stellar disc of the Milky Way, and completely exclude the possibility that it could be an extragalactic source (of any sort: galaxy, supernova, or AGN/LRD). In Section~\ref{sec:data}, we describe the datasets used in our analysis. Then, in Section~\ref{section:proper_motion}, we describe our photometric and proper motion measurements for Capotauro, and in Section~\ref{sec:sed} present the results of the subsequent spectral fitting. We discuss these new results in the context of other recent studies in Section~\ref{section:discussion}, and finally summarise our main conclusions in Section~\ref{section:conclusion}. Throughout, we adopt a $\Lambda$CDM cosmology with $\rm H_{0}=70\,kms^{-1}$, $\Omega_{M}=0.3$ and $\Omega_{\Lambda}=0.7$ and, unless stated explicitly otherwise, all magnitudes are quoted in the AB system \citep{Oke1974,Oke1983}.

\section{DATA}
\label{sec:data}

Our analysis is based on imaging observations of Capotauro obtained with the {\it JWST} \citep{Gardner2006, Gardner2023} Near Infrared Camera \citep[NIRCam;][]{Rieke2003, Rieke2023a} at two distinct epochs.

The first-epoch observations (hereafter epoch 1) were obtained as part of the Cosmic Evolution Early Release Science Survey (CEERS; {\it JWST} Early Release Science program no. 1345, PI Finkelstein; see \citealt{Finkelstein2022, Finkelstein2023, Bagley2023}), which covered a field including Capotauro in December 2022 during its second observing window. This deep imaging comprises the F115W, F150W, F200W, F277W, F356W, F410M and F444W filters, for which we use the v1.0 public data release. Epoch 1 is supplemented by extra, deep F090W imaging taken as part of Cycle 1 GO 2234 (PI Banados) in June 2023. For this observation we employ our own data reduction, produced using the software \textsc{pencil}, a customised version of the \textit{JWST} data reduction pipeline v1.17.1 (D.~Magee et al., in prep.), with the \textit{jwst\_1410.pmap} CRDS context. Note that, although the F090W imaging is taken six months after the rest of the ``epoch 1'' data, Capotauro is undetected in this passband.

The second-epoch (hereafter epoch 2) observations come from two {\it JWST} Cycle-4 programmes which both observed Capotauro in mid-to-late June 2026: MINERVA (GO 7814, PI Muzzin; see \citealt{muzzin2025}) and SPAM (GO 8559; PIs Davis, Larson). Together, these programmes provide new deep medium-band coverage comparable in depth to the CEERS broadbands and, crucially, a $\simeq3.5$-year baseline relative to the earliest observations, providing critical cadence to identify any apparent motion for Capotauro. Both programmes deliver short-wavelength F140M, F162M, F182M and F210M imaging; in the long-wavelength channel, MINERVA provides F250M, F360M and F460M imaging, complemented by SPAM imaging in the F300M, F335M, F430M and F480M passbands. SPAM additionally includes deep broadband F070W imaging. We again use our own \textsc{pencil} reductions for both the MINERVA and SPAM observations, in this case adopting the CRDS context \textit{jwst\_1414.pmap}.

\begin{figure*}
    \centering
    \includegraphics[width=\textwidth]
        {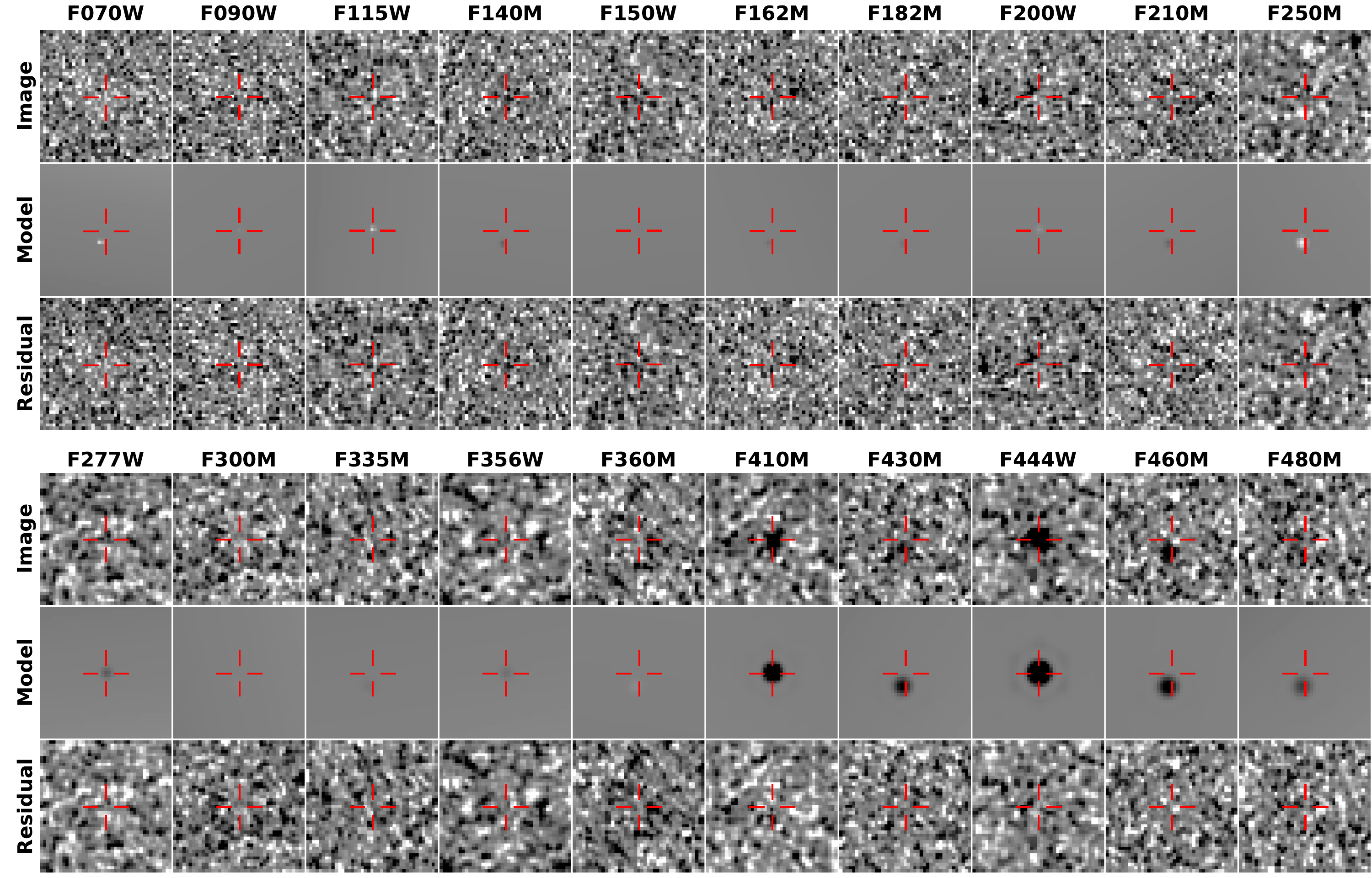}
    \caption{
    {\it JWST}/NIRCam cutouts ($1.2^{\prime\prime}\times1.2^{\prime\prime}$) of Capotauro, ordered by increasing filter wavelength. For each band, the science mosaic is shown at the top, the best-fitting point-source PSF plus local background model in the middle, and the residual at the bottom. The crosshair in every panel marks the source position
    reported by \citet{gandolfi2026} for the December 2022 CEERS observations. The F115W, F150W, F200W, F277W, F356W, F410M, and F444W images were obtained in December 2022, the F090W image was obtained in June 2023, and the F070W, F140M, F162M, F182M, F210M, F250M, F300M, F335M, F360M, F430M, F460M, F480M images were obtained in June 2026. The F430M and F460M observations taken at later times (June 2026) demonstrate a clear movement of Capotauro ($\simeq0.13^{\prime\prime}$) from the first epoch F444W and F410M (December 2022).
    }
    \label{fig:capotauro_jwst_cutouts}
\end{figure*}

The global 5$\sigma$ limiting magnitudes in each filter across the corresponding survey mosaics are presented in Table~\ref{tab:jwst_photometry}. These were measured using 0.2$^{\prime\prime}$-diameter apertures on imaging PSF-homogenized to match the F444W resolution (except F430M, F460M and F480M, which already exhibit a broad flux curve of growth with aperture radius) and corrected to point-source total. We note that these represent the global mosaic detection limits; for the photometry of Capotauro itself, we applied a more thorough local uncertainty measurement. We present all NIRCam cutouts of Capotauro in Fig.~\ref{fig:capotauro_jwst_cutouts}.

\begin{table}
\centering
\caption{{\it JWST}/NIRCam PSF photometry of Capotauro. The first column identifies the filter; the second gives the exposure-time-weighted mean MJD of the calibrated exposures covering Capotauro; the third gives the PSF flux density and its adopted $1\sigma$ uncertainty; the fourth gives the empirical local $2\sigma$-equivalent upper limit for non-detections; the fifth gives the SNR for detections; the sixth gives the survey-wide $5\sigma$ point-source depth in AB mag. The empirical upper limit is defined as the 97.725th percentile of the local blank-position flux distribution after subtraction of its median.}
\label{tab:jwst_photometry}
\footnotesize
\setlength{\tabcolsep}{3pt}
\begin{tabular}{lccccc}
\hline
Filter & $\mathrm{MJD}_{\rm eff}$ & Measured flux &
Upper limit & S/N & $5\sigma$ depth \\
 & (d) & (nJy) & (nJy) & & (AB) \\
\hline
F070W & 61217.4 & $-1.2 \pm 0.8$ & $<1.7$ & -- & 28.7 \\
F090W & 60122.1 & $-0.2 \pm 0.6$ & $<1.2$ & -- & 28.8 \\
F115W & 59936.1 & $-0.6 \pm 0.4$ & $<0.9$ & -- & 29.2 \\
F140M & 61215.5 & $+0.8 \pm 1.0$ & $<2.4$ & -- & 28.6 \\
F150W & 59935.1 & $+0.1 \pm 0.8$ & $<1.6$ & -- & 29.1 \\
F162M & 61215.5 & $+0.5 \pm 1.3$ & $<2.4$ & -- & 28.7 \\
F182M & 61215.2 & $+0.5 \pm 1.3$ & $<2.6$ & -- & 28.8 \\
F200W & 59935.1 & $-0.5 \pm 0.9$ & $<1.9$ & -- & 29.4 \\
F210M & 61215.4 & $+1.3 \pm 1.5$ & $<3.3$ & -- & 28.7 \\
F250M & 61211.9 & $-4.9 \pm 2.3$ & $<5.0$ & -- & 28.3 \\
F277W & 59938.0 & $+0.9 \pm 1.0$ & $<2.0$ & -- & 29.3 \\
F300M & 61217.6 & $-0.6 \pm 2.0$ & $<4.3$ & -- & 28.6 \\
F335M & 61217.6 & $+0.7 \pm 1.9$ & $<4.1$ & -- & 28.7 \\
F356W & 59936.5 & $+0.8 \pm 1.0$ & $<2.1$ & -- & 29.3 \\
F360M & 61217.5 & $-1.2 \pm 1.9$ & $<3.8$ & -- & 28.7 \\
F410M & 59935.0 & $+16.6 \pm 2.6$ & -- & $6.3$ & 28.5 \\
F430M & 61217.4 & $+16.1 \pm 4.4$ & -- & $3.7$ & 27.8 \\
F444W & 59935.1 & $+19.4 \pm 1.8$ & -- & $11.0\phantom{0}$ & 28.8 \\
F460M & 61212.0 & $+32.1 \pm 7.6$ & -- & $4.2$ & 27.5 \\
F480M & 61217.5 & $+16.4 \pm 6.7$ & -- & $2.4$ & 27.6 \\
\hline
\end{tabular}
\end{table}

We also include archival imaging obtained with the \textit{JWST} Mid-Infrared Instrument \citep[MIRI,][]{Glasse2015, Wright2023} as part of the MIRI EGS Galaxy and AGN survey (MEGA; GO 3794, PI Kirkpatrick; see \citealt{Backhaus2025}). These observations were acquired on March 2024 and cover Capotauro in F770W, F1000W, F1500W, and F2100W. We work directly with the MAST public calibrated stage-2 \texttt{cal} exposures, processed with the \textit{JWST} pipeline v1.20.2 \citep{Bushouse2025} and CRDS context \textit{jwst\_1464.pmap}. Two pointings provide seven target-covering exposures in each filter. In one exposure, however, Capotauro falls on invalid detector pixels; we therefore exclude this exposure in all four bands, leaving six usable exposures per filter. The resulting total exposure times are 1665.024\,s in F770W and F1000W, 2031.330\,s in F1500W, and 3080.292\,s in F2100W.

In addition to the datasets described above, Capotauro was also observed as part of the CAPERS survey (GO 6368; PI Dickinson; see \citealt{Donnan2025b}), which obtained \textit{JWST} Near Infrared Spectrograph \citep[NIRSpec;][]{Jakobsen2022, Boker2023} spectroscopy of this target in March 2025. Since the spectrum provides no additional astrometric information on the source, which is the focus of this paper, we do not include it in our analysis; however, we present a brief qualitative discussion in Section~\ref{section:discussion}.

\begin{figure*}
    \centering
    \begin{minipage}[t]{0.335\textwidth}
        \vspace{0pt}\centering
        \includegraphics[width=\linewidth]{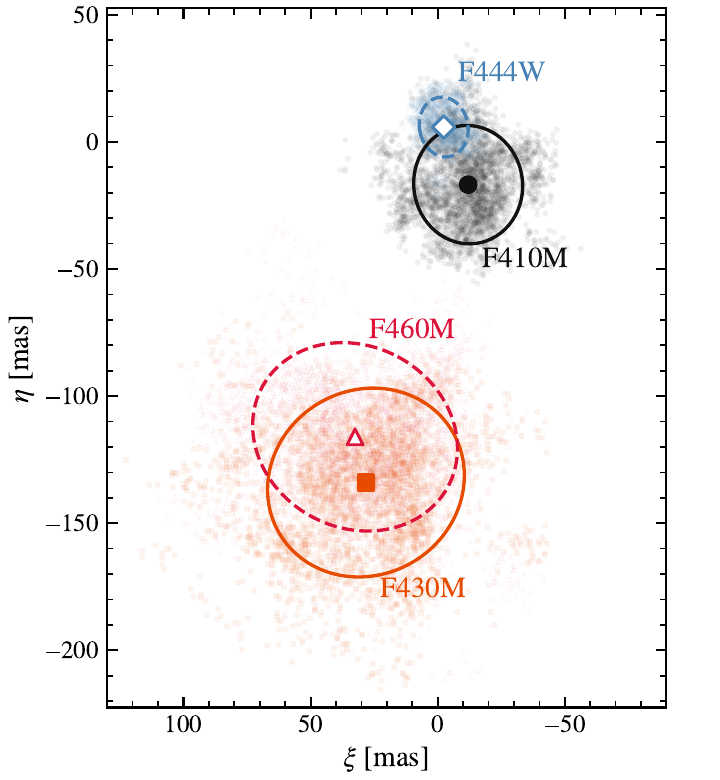}
    \end{minipage}\hfill%
    \begin{minipage}[t]{0.335\textwidth}
        \vspace{0pt}\centering
        \includegraphics[width=\linewidth]{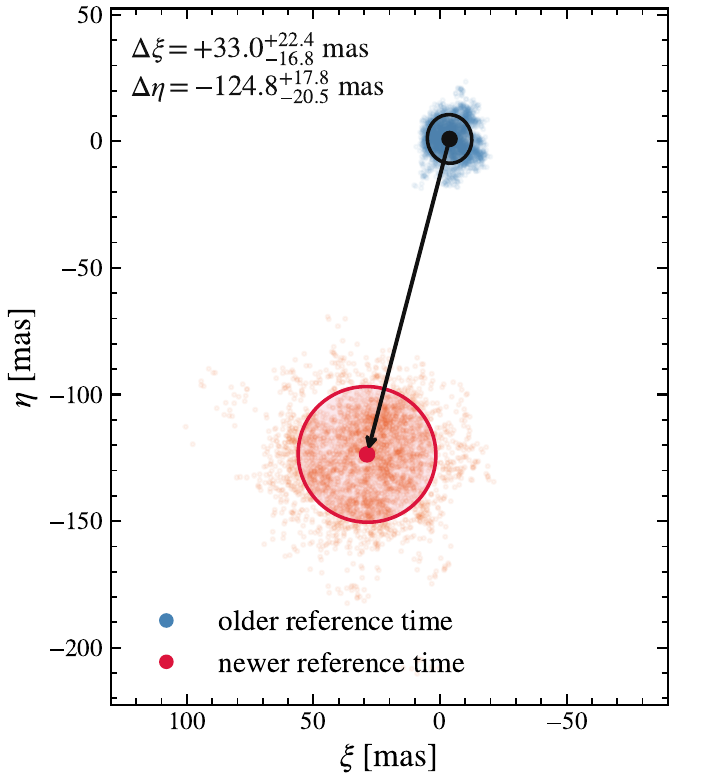}
    \end{minipage}\hfill%
    \begin{minipage}[t]{0.325\textwidth}
        \vspace{0pt}\centering
        \includegraphics[width=\linewidth]{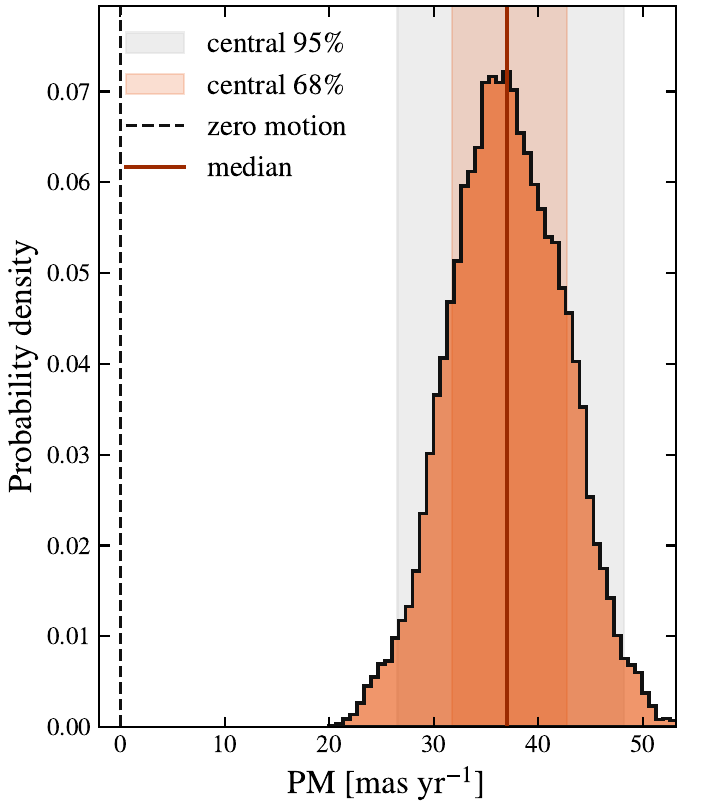}
    \end{minipage}
    \caption{Relative astrometry of Capotauro using the corrected target-covering exposure times. \emph{Left:} independently measured positions in each of the two filters from epoch 1 (December 2022), F410M and F444W (black and blue symbols), and the two filters from epoch 2 (June 2026), F430M and F460M (red and orange symbols). Points show representative subsets of the propagated MC draws and outlines enclose 68\% of each image's two-dimensional distribution. \emph{Middle:} the proper-motion-only model evaluated at representative times of the epoch-1 and -2 observing groups as trajectory-evaluation clouds, and the arrow shows the inferred displacement. The text gives the eastward and northward offsets with central 68\% intervals. \emph{Right:} empirical distribution of each two-dimensional motion draw projected onto the median motion direction. Zero is the stationary-source hypothesis, while the orange and grey regions show central 68\% and 95\% intervals. North is upwards and positive east is towards the left in the positional panels.}
    \label{fig:capotauro_positional_offset}
\end{figure*}

\section{Photometric and astrometric measurements}
\label{section:proper_motion}

In this section, we describe how we measure both the astrometric positions and the photometric fluxes of Capotauro. In both cases we apply several independent techniques to the same data, in order to test whether the derived values are robust to the choice of method. We first present the astrometry, which is based on fitting a model of the point spread function (PSF) to each image and yields the trajectory of the source, and then the photometry, which uses the same PSF fits and relies on this trajectory to place its forced measurements.

\subsection{Positional offset measurement}\label{sec:offset_measurement}

Capotauro is unresolved in all of the NIRCam images (Fig.~\ref{fig:capotauro_jwst_cutouts}), but visibly appears to show clear movement between the first and second epochs (Fig.~\ref{fig:capotauro_positional_offset}). We therefore measured its properties in each image by fitting a PSF model, which yields both the flux and the centroid of the source (Section \ref{sec:position_1}). Next, we considered and modelled all potential sources of positional uncertainty (Section \ref{sec:position_2}). Finally, we constrained the apparent motion of Capotauro (Section \ref{sec:position_3}).

\subsubsection{Centroid measurement} \label{sec:position_1}
To model the centroid of Capotauro in all four NIRCam images, we generated a filter-dependent, polychromatic NIRCam PSF model for each band using \texttt{STPSF} v2.2.0\footnote{See details on the {\it JWST} PSF Simulation Tool:\\ \href{https://www.stsci.edu/jwst/science-planning/proposal-planning-toolbox/psf-simulation-tool}{https://www.stsci.edu/jwst/science-planning/proposal-planning-toolbox/psf-simulation-tool}} \citep{Perrin2012, Perrin2014}. Nine monochromatic PSFs were computed across each medium-band filter passband and 21 across each wide-band passband, and were combined with weights given by the photon-count distribution expected for a cool brown dwarf, represented by an \texttt{ATMO}~2020 chemical-equilibrium atmosphere model with $\Teff=300$\,K and $\log{g}=4.5$\,dex \citep{Phillips2020}. This template is representative of the cool dwarfs typically found in deep extragalactic fields \citep[e.g.,][]{hainline2026b} and approximates the spectral energy distribution (SED) that we eventually derive for Capotauro (Section~\ref{sec:sed}). We verified that this choice has a negligible influence on the final results by replacing this template with a 300-K blackbody SED and a flat SED. The oversampling factors for the detector plane and the Fourier propagation were both set to four. The resulting models were resampled to the $0.03''$ pixel scale of the mosaics and normalized so that the pixel sum of each template is unity, and the fitted PSF amplitude therefore corresponds directly to the integrated source flux. The same PSF models are used for the photometry in Section~\ref{sec:photom_measurement}.

We use the fitted centroids in F410M, F444W, F430M, and F460M as four independent position measurements, expressed in a local tangent plane with positive $\xi$ eastwards and positive $\eta$ northwards. We retained the celestial WCS propagated from the calibrated \textit{JWST} products. The WCS roll uncertainty is negligible and we corrected the relative offsets in Section~\ref{sec:position_2}. In epoch 1, Capotauro is strongly detected in F410M and F444W (with signal-to-noise ratios above 6; see Section~\ref{sec:photom_measurement}), and we fitted the amplitude and centroid simultaneously in each image, taking the coordinates reported by \citet{gandolfi2026} as the initial estimate and allowing the centroid to vary within $0.18''$ of them. The exposure-weighted mean epochs of the two images are MJD 59935.019 and 59935.082, separated by 1.5 h. These therefore provide independent centroid measurements but sample a single parallax phase. Indeed, their parallax factors, which describe the apparent displacement that the orbital motion of the observatory would induce for a source at finite distance, are effectively identical: $(P_\xi,P_\eta)=(+0.754,-0.649)$ and $(+0.754,-0.648)$. We computed these factors from the actual position of {\it JWST} about the Sun-Earth $L_2$ point \citep{Gardner2023}, sampling the reconstructed spacecraft trajectory for every exposure from the JPL \textsc{Horizons} system (spacecraft ID $-170$; \citealt{Giorgini1996}) as the geometric ICRF equatorial Cartesian position relative to the Solar-system barycentre; the spacecraft-ephemeris uncertainty is negligible.

In epoch 2, we repeated a blind, positive-amplitude PSF-centroid search in F430M and F460M, independently in each image, within $0.65''$ of the predicted position. The derived peak locations in each image agree within 90 mas and are similarly retained as independent astrometric measurements. Their target-covering mean epochs are MJD 61217.435 and 61211.977, separated by 5.46 d, with parallax factors $(-0.818,+0.629)$ and $(-0.767,+0.691)$.

\subsubsection{Positional uncertainty} \label{sec:position_2}
\begin{figure*}
    \centering
    \begin{minipage}[t]{0.58\textwidth}
        \vspace{0pt}
        \centering
        \includegraphics[width=\linewidth]{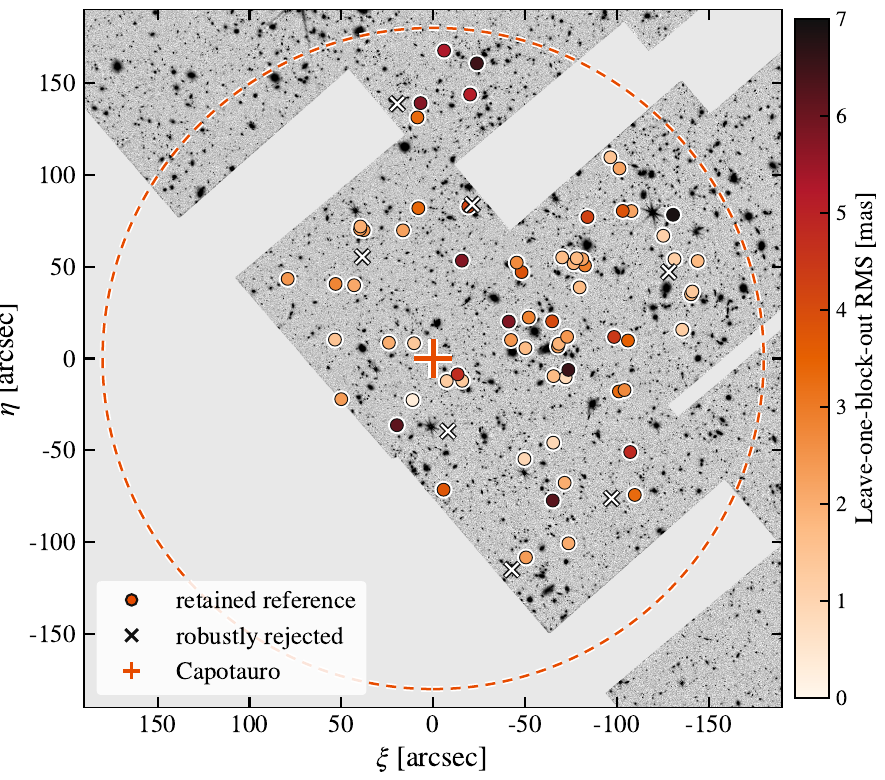}
    \end{minipage}\hfill%
    \begin{minipage}[t]{0.41\textwidth}
        \vspace{0pt}
        \centering
        \includegraphics[width=\linewidth]{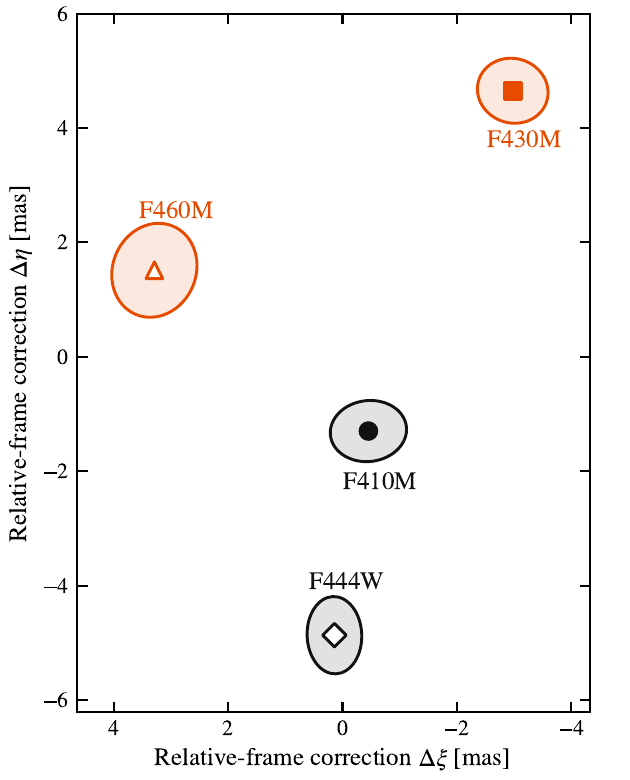}
    \end{minipage}
    \caption{Validation of the local relative-astrometric reference frame. \emph{Left:} F444W image of the common four-band field surrounding Capotauro. The orange cross marks Capotauro and the dashed circle has a radius of $3'$. Circles identify compact reference sources retained in the alignment, with colour giving the radial RMS when the corresponding $30''$ spatial block is omitted from the four-image solution; crosses mark robustly rejected sources. This leave-one-block-out test measures prediction stability in regions excluded from a fit. \emph{Right:} relative-frame corrections at Capotauro for the four astrometric images. Black symbols denote F410M and F444W, while orange symbols denote F430M and F460M; marker shapes identify filters. Ellipses enclose 68\% of the adopted two-dimensional frame-correction distribution. North is upwards and positive east is towards the left in both panels.}
    \label{fig:astrometric_frame_validation}
\end{figure*}

We consider all major sources of positional uncertainty. The first is the astrometric misalignment between the four mosaics, which we measured using reference sources in the common four-image footprint within $3'$ of Capotauro (Fig.~\ref{fig:astrometric_frame_validation}). After robust quality selection, 68 isolated, bright, compact, and predominantly galaxy-like reference sources remained. We fitted their four-band centroids simultaneously, with one latent position per reference and one relative affine transformation per image, under the convention that the four frame corrections sum to zero. This uses all six combinations among the four images and requires no tie to an absolute reference frame, because a translation common to all images cancels from the motion of Capotauro relative to the field. While frames have relative and correctable offsets, the measurement of these expected corrections still carries uncertainties. The correlated frame uncertainty was estimated with 5,000 bootstrap resamples over 36 spatial blocks, each $30''$ across, and fits that excluded one block at a time verified that the solution predicts positions reliably in regions it was not fitted to (Fig.~\ref{fig:astrometric_frame_validation}). Because galaxy morphology can vary coherently with wavelength and mimic a frame shift, we also formed 34 spatial pairs of references, repeated the solution after 12 splits in brightness, colour, size, and axis ratio, and evaluated each diagnostic with 10,000 within-pair label permutations. No galaxy morphological impact remained significant, and we record a residual uncertainty of only 0.1 to 0.3 mas as a four-image correlated covariance for other later apparent motion fitting. At the position of Capotauro, the reference frame corrections in F410M, F430M, F444W, and F460M are $(-0.5,-1.3)$, $(-3.0,+4.7)$, $(+0.1,-4.9)$, and $(+3.3,+1.5)$ mas, respectively (Fig.~\ref{fig:astrometric_frame_validation}). The frame uncertainties after correction between filters spanning the two epochs are negligible, e.g., only $(0.8,0.6)$ mas between F410M and F430M and $(0.8,0.9)$ mas between F444W and F460M.

A larger positional uncertainty comes from the structured noise underneath Capotauro, which can displace a fitted PSF centroid even if the image alignment is exact. We quantified this effect with a joint four-image injection-recovery Monte Carlo (MC) simulation; the same suite of artificial-source injections also calibrates the photometric uncertainties and biases in Section~\ref{sec:photom_measurement}. Artificial moving point sources were placed at 75 source-masked background locations within $20''$ of Capotauro, separated by at least $2.7''$, and each location was sampled on a $4\times4$ sub-pixel phase grid, producing 1,200 physical four-image trials. Inserting the artificial PSFs into source-screened patches of the real mosaics preserves the correlated noise, large-scale background structure, and crowding of the data. The planted trajectories followed the converged proper-motion-only solution evaluated at the corrected local time of each image, while the input flux at each location was conditioned on the PSF amplitude measured from that same blank patch, and recovery repeated the complete estimator used for Capotauro, including the centroid search, source association, and selection. We then performed 5,000 bootstrap resamples over whole background locations, which preserves the correlations among the four images and among the 16 sub-pixel phases and propagates the uncertainty from the finite number of background patches into the median centroid-bias calibration. For each image, the adopted direct position is the measured centroid plus the relative-frame correction and minus the median recovered-minus-injected residual. The final positional distribution combines the empirical centroid errors, the finite bias-calibration uncertainty, the frame-bootstrap covariance, and the reference-morphology covariance. After the corrected trajectory was re-injected, all four position updates were smaller than 1.6 mas, confirming convergence. The marginal standard deviations in $(\xi,\eta)$ are $(14.2,15.4)$ mas for F410M, $(25.6,24.6)$ mas for F430M, $(6.4,7.8)$ mas for F444W, and $(26.7,24.6)$ mas for F460M; Table~\ref{tab:individual_astrometry} gives the asymmetric one-dimensional intervals and the radii enclosing 68\% of the two-dimensional distributions.

\begin{table}
\centering
\caption{Individual-image relative astrometry of Capotauro. The first column identifies the filter; the second and third give the raw measured PSF centroid; the fourth and fifth give the adopted reference-frame-corrected coordinates with central 68\% marginal intervals; the final column gives the radius enclosing 68\% of the two-dimensional positional distribution, as illustrated in Fig.~\ref{fig:capotauro_positional_offset}. Coordinates are measured in the tangent plane centred on $(\alpha_0,\delta_0)=(214.887376^\circ,52.797809^\circ)$, with positive $\xi$ eastwards and positive $\eta$ northwards.}
\label{tab:individual_astrometry}
\setlength{\tabcolsep}{4.5pt}
\renewcommand{\arraystretch}{1.42}
\begin{tabular}{lccccc}
\hline
Filter & $\xi_{\rm raw}$ & $\eta_{\rm raw}$ & $\xi_{\rm dir}$ & $\eta_{\rm dir}$ & $r_{68}$ \\
 & (mas) & (mas) & (mas) & (mas) & (mas) \\
\hline
F410M  & $-10.8$ & $-16.1$ & $-12.0_{-11.4}^{+15.3}$ & $-16.8_{-14.3}^{+16.1}$ & $23.1$ \\
F444W & $-3.5$ & $+9.4$ & $-2.4_{-6.8}^{+6.5}$ & $+5.9_{-6.4}^{+8.7}$ & $10.8$ \\
F460M & $+29.1$ & $-121.8$ & $+32.5_{-27.6}^{+28.6}$ & $-116.0_{-24.4}^{+20.9}$ & $37.6$ \\
F430M & $+33.9$ & $-144.4$ & $+28.2_{-20.5}^{+28.3}$ & $-134.0_{-28.3}^{+20.9}$ & $37.9$ \\
\hline
\end{tabular}
\end{table}

\subsubsection{The apparent motion of Capotauro} \label{sec:position_3}
We fitted all four direct positions simultaneously at their individual local epochs. Our primary model is a linear two-dimensional trajectory with the parallax fixed to zero, which directly describes the apparent motion sampled by these data. The wavelength-matched pairs, F410M with F430M and F444W with F460M, independently give median displacements of $(+39.6,-120.2)$ and $(+35.7,-122.3)$ mas, demonstrating that the motion is present in both image pairs, and combining their largely independent centroid errors increases the precision. The band-mean baseline from epoch 1 to epoch 2 is 3.5035 yr. The stationary and linear-motion fits give $\chi^2/\nu=44.4/6$ and $2.1/4$, respectively, so the motion is preferred with $\Delta\chi^2=42.3$ for two additional parameters, corresponding to a two-sided Gaussian-equivalent significance of $6.2\sigma$ (Fig.~\ref{fig:capotauro_positional_offset}). The apparent-motion solution is $(\mu_\xi,\mu_\eta)=(+9.4^{+6.4}_{-4.8},-35.6^{+5.1}_{-5.9})$ mas yr$^{-1}$, with a total motion of $37.6^{+5.5}_{-5.6}$ mas yr$^{-1}$ at a position angle of $165.3^{+7.2}_{-9.6}$ deg east of north. This motion also consistent with the upper limit of \citet{gandolfi2026b}, whose 2.3-year baseline corresponds to a predicted displacement of $\sim$90 mas for our solution, below their limit of 137 mas.

For completeness, we also allowed a parallax to vary together with a proper motion component, using the fixed {\it JWST} parallax factors quoted above and no positivity or distance prior. Because the two filters in each epoch sample almost the same parallax phase, the data provide effectively only two astrometric epochs, and the degeneracy between proper motion and parallax is nearly complete. This diagnostic fit gives $\varpi=122^{+445}_{-434}$ mas, with a central 95\% interval of $[-720,+933]$ mas, and improves the fit by only $\Delta\chi^2=0.08$ for one extra parameter ($0.3\sigma$). We therefore report a non-detection of parallax with the current data, and we summarize our findings in Table~\ref{tab:astrometric_models}.

As a sensitivity test for any mismatch between the simulated and effective mosaic PSFs, we derived an empirical broadening correction independently in each astrometric band. \textit{Gaia} DR3 \citep{2016A&A...595A...1G, 2023A&A...674A...1G} was used to identify likely point sources with $17<G<21$ and ${\rm RUWE}<1.4$. We convolved the nominal \texttt{STPSF} template with an exactly centred circular Gaussian and fitted its additional width, the PSF amplitude, and a tilted background for each usable star. A 1,000-realization star-level bootstrap of the 13, 14, 13, and 13 accepted stars in F410M, F430M, F444W, and F460M yielded additional kernel widths of $33.0\pm0.8$, $30.8\pm0.8$, $30.0\pm0.8$, and $30.0^{+0.8}_{-0.1}$ mas, respectively. Repeating the Capotauro fits with these broadened templates changes the final apparent displacement between the two epochs by only $(-0.4,-0.1)$ mas in $(\xi,\eta)$, demonstrating that the motion detection is insensitive to the effective PSF width.

As a further complementary test at the reduction level, we repeated the astrometric analysis using the 29 individual stage-2 NIRCam \texttt{cal} exposures retrieved from MAST, rather than the SWarp/LANCZOS3-resampled mosaics. Working on the native detector grids avoids interpolation-induced correlated noise and preserves the detector position, dither, data-quality mask, and exact midpoint time of every exposure; it also permits a detector-position- and date-dependent \texttt{STPSF} model and an exposure-specific {\it JWST} parallax factor. The corresponding disadvantage is that Capotauro is detected less significantly in any one native exposure than in the combined mosaics, so fitting 29 unconstrained centroids would allow the weakest exposures to follow local noise peaks. We therefore fitted one common sky trajectory directly to all native pixels, while profiling one non-negative source amplitude per filter and an independently tilted background in each exposure. The local relative frame was constructed from 287 stable reference sources represented by 3,388 detections, with its uncertainty propagated through 5,000 spatial-block bootstrap realizations, and the displacement caused by structured background noise was calibrated with 1,000 coherent 29-exposure injection-recovery experiments followed by 5,000 spatial-cluster bootstrap realizations. The native-exposure fit gives $(\mu_\xi,\mu_\eta)=(+12.8^{+5.8}_{-4.8},-37.8^{+3.7}_{-6.0})$ mas yr$^{-1}$, corresponding to a total motion of $40.8^{+4.9}_{-4.3}$ mas yr$^{-1}$ at a position angle of $161.7^{+6.4}_{-8.4}$ deg east of north, and its stationary-versus-moving likelihood ratio corresponds to $7.6\sigma$. These values differ from the primary mosaic result by only $(+3.4,-2.1)$ mas yr$^{-1}$ in the two components and by 3.3 mas yr$^{-1}$ in total motion, all well within the empirical 68\% intervals. The two reductions are therefore highly consistent, showing that the inferred motion is produced neither by the mosaic resampling nor by the effective PSF of the mosaics. Allowing a signed parallax in the native fit gives a nominal $\varpi=-51.9$ mas, but injections of zero-parallax sources give $p=0.25$ and the physically constrained solution lies at $\varpi=0$; this cross-check likewise provides no positive-parallax detection or astrometric distance.

\begin{table}
\centering
\caption{Capotauro motion diagnostics. Uncertainties are central 68\% intervals of the empirical MC distributions.}
\label{tab:astrometric_models}
\setlength{\tabcolsep}{5pt}
\renewcommand{\arraystretch}{1.28}
\begin{tabular}{@{}lcc@{}}
\hline
Quantity & Unit & Result \\
\hline
$\Delta\xi_{\rm app}$ & mas & $+33.0_{-16.8}^{+22.4}$ \\
$\Delta\eta_{\rm app}$ & mas & $-124.8_{-20.5}^{+17.8}$ \\
$\Delta r_{\rm app}$ & mas & $131.7_{-19.6}^{+19.1}$ \\
\hline
$\mu_\xi$ & mas yr$^{-1}$ & $+9.4_{-4.8}^{+6.4}$ \\
$\mu_\eta$ & mas yr$^{-1}$ & $-35.6_{-5.9}^{+5.1}$ \\
$\mu$ & mas yr$^{-1}$ & $37.6_{-5.6}^{+5.5}$ \\
${\rm PA}_\mu$ & deg & $165.3_{-9.6}^{+7.2}$ \\
$\chi^2/\nu$ & --- & $2.1/4$ \\
$\Delta\chi^2_{\rm motion}$ & --- & $42.3$ ($6.18\sigma$) \\
\hline
$\varpi_{\rm free}$ & mas & $122_{-434}^{+445}$ \\
$\chi^2/\nu$ & --- & $2.00/3$ \\
$\Delta\chi^2_{\varpi}$ & --- & $0.077$ ($0.28\sigma$) \\
\hline
\end{tabular}
\begin{minipage}{0.96\columnwidth}
\footnotesize\textit{Notes.} Position angles are measured east of north.
\end{minipage}
\end{table}

\subsection{Photometric measurement}
\label{sec:photom_measurement}

The PSF fits described in Section~\ref{sec:offset_measurement} also provide the photometry of Capotauro. For a faint point source, PSF fitting is less affected by the noise structure underneath the source than fixed-aperture photometry, which is critical for an accurate flux measurement of a faint source. Because Capotauro moves between the two epochs, we fitted the images of each epoch separately. The fluxes in the four astrometric bands come directly from the fits of Section~\ref{sec:offset_measurement}: the free fits in F410M and F444W in epoch 1, which yield detections with signal-to-noise ratio ${\rm SNR}>6$ (Table~\ref{tab:jwst_photometry}), and the independent searches in F430M and F460M in epoch 2. In all other images, we performed forced PSF photometry, fixing the position to that predicted by the astrometric trajectory at the time of each observation; this prevents the fit from settling on a nearby noise fluctuation when the source is not detected.

A locally tilted background was fitted simultaneously with the PSF in every image. To characterize residual background structure and correlated mosaic noise, we repeated the identical PSF-plus-background measurement at 1,000 source-masked trial positions within $20''$ of Capotauro. Detected objects and their surroundings were excluded, with larger masks applied to bright structures, and a trial was rejected whenever a masked pixel entered the fitting region. The median of the blank-position amplitudes was used to correct the small local offset of the flux zero point, while their robust dispersion provided the background-derived $1\sigma$ flux uncertainty.
The photometric performance of this procedure was tested with the same artificial-source injections introduced in Section~\ref{sec:offset_measurement}. In each of the four detection bands, F410M, F430M, F444W, and F460M, we conducted 1,000 trials at the measured flux, supplemented by 250 trials each at fluxes one background-derived standard deviation below and above that value, with recovery again repeating the same centroid search, source association, selection, and PSF fit used for the main measurement. For each band, we conservatively adopted the larger of the blank-position dispersion and the injection-recovery flux scatter as the final random uncertainty. The injection scatter increased the adopted uncertainty in F430M, F444W, and F460M, whereas the blank-derived value remained slightly larger in F410M. The simulations also quantified the flux bias introduced by searching for and selecting a positive peak, which was negligible in all bands, with the largest value occurring in F430M. Using a detection threshold of $2\sigma$, Capotauro is detected in F410M, F430M, F444W, F460M, and F480M (Table~\ref{tab:jwst_photometry}). The remaining NIRCam bands are reported as non-detections with their empirical $2\sigma$ upper limits.

As a verification, we also performed alternative photometry for Capotauro using \textsc{Source-Extractor} \citep{1996A&AS..117..393B} in dual-image mode, creating two detection catalogues to account for the positional shift of Capotauro between the epochs: a F444W-detected catalogue for all epoch 1 measurements (the CEERS broad bands, plus F410M and the GO 2234 F090W imaging), and a F460M-detected catalogue for all of the later MINERVA+SPAM measurements. We used $0.2''$-diameter apertures on imaging that was PSF-homogenized to F444W, with the exception of the F430M, F460M, and F480M imaging that already had a broad PSF. To measure photometric uncertainties, we follow the method described in our previous works (e.g., \citealt{begley2025}), whereby we measure the aperture-to-aperture rms of the nearest 200 blank-sky $0.2''$ apertures after masking sources with a segmentation map. All measurements were then corrected to point-source total fluxes. Encouragingly, the two methods return fully consistent flux measurements across all bands, and we adopt the PSF-fitting measurements as fiducial for the remainder of this work.

We note that our photometry differs from that reported by \citet{gandolfi2026}, by as much as $\simeq58\%$ when comparing to the PSF-fitting method, and $\simeq44\%$ if using the point-source-corrected $0.2''$ apertures. This difference has important implications for the potential variability discussed by \citet{ferrara2026} (see Section~\ref{section:variability}), and so warrants investigation. As \citet{gandolfi2026} adopt Kron aperture photometry \citep{Kron1980}, we checked the earlier CEERS v0.5 and v0.6 data release imaging (i.e., that utilised by \citealt{gandolfi2026}) and find that adopting Kron apertures on the F444W imaging reproduces a very similar flux (29.6 nJy, versus the $30.7\pm2.4$ nJy reported by \citealt{gandolfi2026}, well within the $1\sigma$ uncertainty). However, the Kron elliptical aperture is erroneously large ($\sim0.69''\times0.60''$ in diameter, $\sim5\times$ larger than NIRCam long-wavelength resolution) relative to the point-like morphology of Capotauro, biased towards nearby overlapped noise, which suggests that their flux measurement may be overestimated.

Finally, we measured the MIRI photometry directly from the six valid exposures in each filter. We generated MIRI PSFs with \texttt{STPSF} v2.2.0 using the same settings as for NIRCam, while additionally including the MIRI detector-scattering and distortion model. In each filter, all six exposures were fitted simultaneously with one common point-source amplitude and an independently tilted local background for every exposure. These were forced measurements at the position predicted for the MIRI observing time. We characterized the local noise by repeating the complete six-exposure fit at 1,800 source-masked positions within $20''$ of Capotauro, of which 217 satisfied the same exposure-coverage, valid-pixel, and detector-edge requirements as the target. We additionally performed 1,000 injection-recovery trials per filter, including a 35-mas positional perturbation, and adopted the larger of the blank-position and injection-recovery dispersions as the random uncertainty. Five thousand spatial-block bootstrap realizations were then used to propagate the finite number of independent background regions and to determine the empirical covariance between the four filters, while a separate 5\% calibration uncertainty is retained for the model fitting. Capotauro is not detected in any MIRI filter, and we report the MIRI photometry in Table~\ref{tab:miri_photometry}, together with $2\sigma$-equivalent upper limits.

\begin{table}
\centering
\caption{{\it JWST} MIRI forced PSF photometry of Capotauro. The first column identifies the filter; the second gives the signed fitted flux density and empirical $1\sigma$ uncertainty used in the SED likelihood; the third gives the 2-$\sigma$-equivalent upper limit; the fourth gives the corresponding lower limit on AB mag..}
\label{tab:miri_photometry}
\small
\begin{tabular}{lccc}
\hline
Filter & Measured flux & Upper limit & AB limit \\
 & (nJy) & (nJy) & \\
\hline
F770W  & $-5.8\pm35.2$     & $<83.6$   & $>26.6$ \\
F1000W & $+83.0\pm79.4$    & $<173.0$  & $>25.8$ \\
F1500W & $-139.0\pm188.7$  & $<450.5$  & $>24.8$ \\
F2100W & $-1230.8\pm889.9$ & $<2120.1$ & $>23.1$ \\
\hline
\end{tabular}
\end{table}

\begin{figure*}
    \centering
    \includegraphics[width=0.9\textwidth]{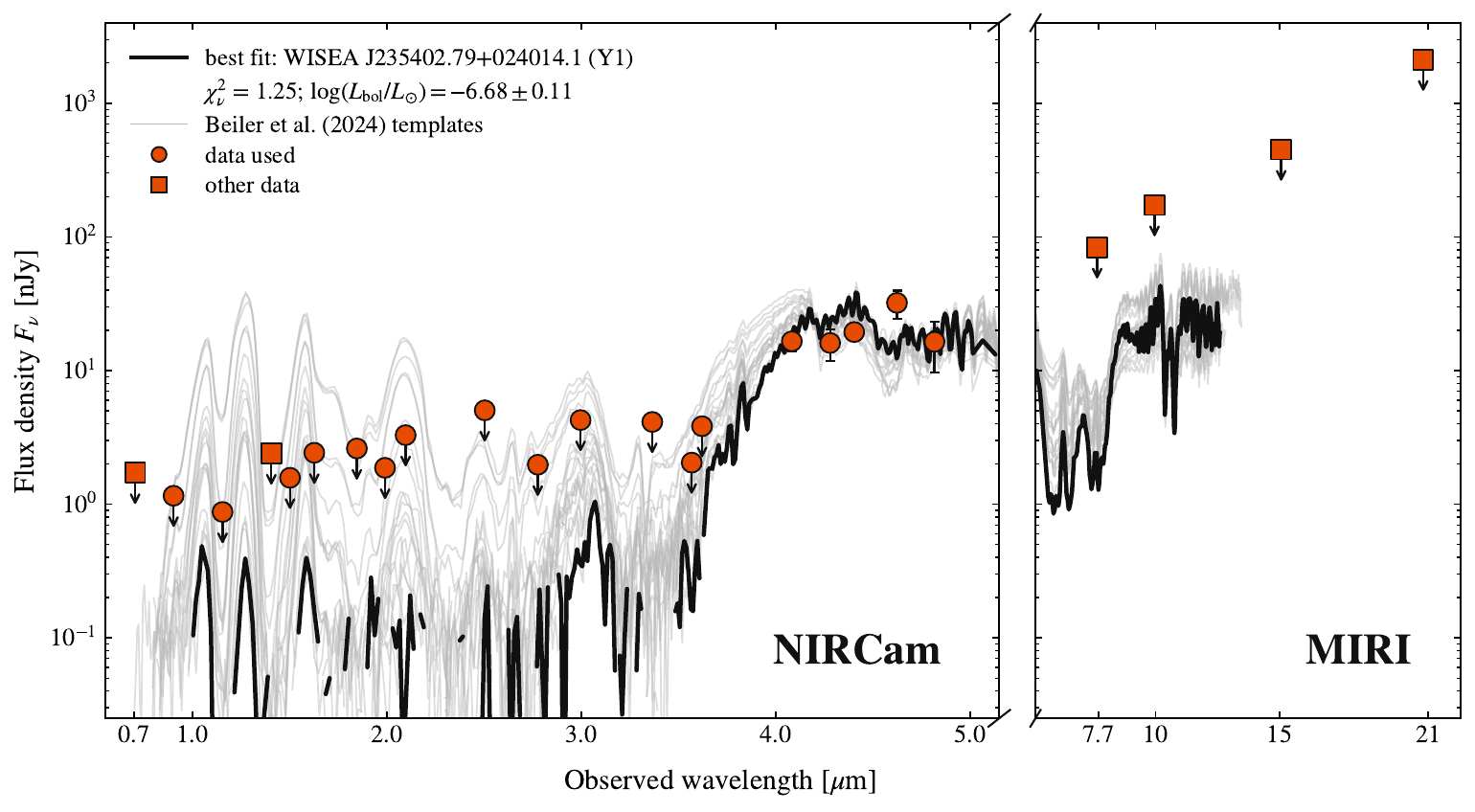}
    \vspace{-0.1in}
    \caption{
    SED of Capotauro compared with the empirical {\it JWST} late-T and Y~dwarf templates from \citet{Beiler2024}. 
    Grey curves show the template library, while the black curve shows the best-matching template, the Y1~dwarf WISEA~J235402.79+024014.1. 
    Orange circles indicate photometric measurements used in the $\chi^2$ fit, and orange squares show additional photometry not included in the fit. 
    The comparison yields a reduced $\chi_\nu^2=1.25$ and is consistent with a Y-dwarf SED with an inferred luminosity of 
    $\log(\Lbol/\Lsun)=-6.68\pm0.11$\,dex.
    }
    \label{fig:capotauro_sed}
\end{figure*}

\section{Spectral energy distribution of Capotauro}
\label{sec:sed}

The apparent motion of Capotauro indicates that it is a Galactic object, and so to further characterise it we compared our photometry to data for other brown dwarfs observed by {\it JWST}. With robust detections in only five filters, spanning 3.9--5.0\,\micron, and with a limited sample of templates, we do not expect to precisely constrain the physical properties of Capotauro. However, we are able to provide estimates of its properties to place it in the context of other known brown dwarfs. Given the unknown distance to Capotauro, we examine only flux ratios in our SED analysis, using the highest S/N flux measurement from F444W as the denominator. This allows us to include all observed bandpasses in our analysis, even the the ones that provide nondetections.

There is only one collection of photometry for brown dwarfs suitable for our analysis, and that is the sample of 23 late-T and Y~dwarfs reported by \citet{Beiler2024}. It includes homogeneous {\it JWST} spectra spanning $\sim$1--12\,\micron\ from which they compute synthetic photometry in all {\it JWST} bandpasses that fall entirely within their observed wavelength range. This covers all filters in our analysis except for F070W, which we therefore exclude here. The only other NIRCam filter we exclude is F140M because several of the colder brown dwarfs in the template sample lack data here, and we want our tests to be as homogeneous as possible. We also choose to exclude MIRI photometry because the upper limits are a few to $\sim$10$\times$ shallower than the NIRCam depths and thus provide no meaningful constraints (all templates would agree given the measured F444W flux). We converted the Vega magnitudes reported in \citet{Beiler2024} to fluxes using the appropriate zero points and then computed flux ratios relative to F444W for all template objects. Given the high-S/N spectra on which the synthetic photometry is based, we neglect any errors associated with the template flux ratios.

The complete sample of templates comprises 20 objects with spectral types ranging from T6--Y1; three of the objects in \citet{Beiler2024} lack data in all 18 bandpasses used for the flux ratios. We excluded one more anomalous object (WISE~J053516.80$-$750024.9) from our analysis because it is a suspected unresolved binary \citep{Leggett2021}, so our final template library includes 19 objects. For each template we computed $\chi^2$, and the best-matching template was the Y1~dwarf WISEA~J235402.79+024014.1 ($\Teff\simeq350$\,K), with $\chi^2=21.3$ for 17 degrees of freedom. The fitted SED is presented in Fig.~\ref{fig:capotauro_sed}. According to the $\chi^2$ distribution, the probability of obtaining a value of $\chi^2$ this high or higher is 0.21, so our flux measurements are indeed well-matched to a Y-dwarf SED. We computed the probability $p_i$ associated with each template $\chi_i^2$ and used these probabilities as weights $w_i \equiv p_i$ to compute the range of properties consistent with our photometry of Capotauro. We computed a mean $\mu$ and standard deviation $\sigma$ in each property $X$ by, respectively, summing over the template properties $\mu = \sum{w_i X_i}/\sum{w_i}$ and $\sigma^2 = \sum{w_i (X_i-\mu)^2}/\sum{w_i}$. This procedure is most useful for properly tracking errors on continuous parameters like luminosity, and our estimate for Capotauro is $\log(\Lbol/\Lsun) = -6.68\pm0.11$\,dex. The resulting $\sigma$ here is $\gtrsim4\times$ the uncertainties on the template values, which justifies our neglecting them in our analysis. We also examined spectral type and find that, to within a half-subtype, the estimate for Capotauro is Y$1.0\pm0.5$. The temperature of the best-matching object ($\simeq350$\,K) is fully consistent with that derived by \citet{hainline2026b} for Capotauro, who found a value of $\Teff=342$\raisebox{0.5ex}{\tiny$\substack{+67 \\ -46}$}\,K having fitted the December 2022 NIRCam and March 2024 MIRI photometry from \citet{gandolfi2026} with Sonora Elf Owl substellar atmospheric models \citep{Mukherjee2024, Mukherjee2025, Wogan2025}, despite the likely overestimation of the \citet{gandolfi2026} NIRCam photometry (see Section~\ref{sec:photom_measurement}).

Before proceeding, we will note here that our analysis is fundamentally limited by the empirical templates available to us. Comparing to theoretical photometry of brown dwarfs would alleviate this issue; however, with photometry covering only a small part of the SED, it would be very difficult or impossible for our analysis to disentangle the many relevant parameters like effective temperature, surface gravity, metallicity, disequilibrium chemistry, carbon-to-oxygen ratio, clouds/rainout \citep[e.g.,][]{Mukherjee2024,Leggett2025}. Moreover, such theoretical models have also not been fully validated by observations and thereby have unknown accuracy. This is why we have chosen a conservative approach that focuses on empirical data for brown dwarfs with robustly known properties. 

The fact that our template sample lacks objects with spectral types later than Y1 means that Capotauro could be even colder than our analysis finds. As an example, the coldest known brown dwarf is WISE~J085510.83$-$071442.5 (Y4 or later; \citealp{Kirkpatrick2019}), and {\it JWST} photometry from \citet{Albert2025} shows that it has a color of $\mathrm{F150W-F480M} = 10.30\pm0.10$\,mag (Vega). Capotauro is barely detected at F480M, and the F150W imaging is 4--5\,mag too shallow to detect it if it were as red as WISE~J085510.83$-$071442.5. Therefore, broadly speaking, the existing imaging data for Capotauro is too shallow to robustly test whether it might be colder than Y1. Given these impediments, our analysis formally provides only limits on the properties of Capotauro, since our best-fit templates are at the edge of the parameter space probed by existing data.

Finally, we can obtain a distance estimate for Capotauro assuming that it is similar to the brown dwarfs in our template sample. The distances to the templates are known from their parallaxes, and we computed the distance for Capotauro needed to make its F444W flux equal to each template's F444W flux. The $\chi^2$-weighted distance we find is $730\pm110$\,pc, where the uncertainty is driven by the dispersion among templates, similarly to the luminosity above. The input measurement errors due to template parallaxes and Capotauro's F444W flux are negligible. This distance is also consistent within 1$\sigma$ with that derived by \citet{hainline2026b}, who found a value of $600$\raisebox{0.5ex}{\tiny$\substack{+280 \\ -180}$}~pc.

\section{Discussion}
\label{section:discussion}

The significant apparent motion of Capotauro, together with its Y-dwarf-like SED confirms that it is a Galactic brown dwarf rather than an extragalactic source. In this section we first use the SED-inferred distance to separate the proper-motion and parallax contributions to the measured displacement (Section~\ref{subsec:parallax&PM}), and then test the available imaging for evidence of variability (Section~\ref{section:variability}). We subsequently discuss the implications of Capotauro for brown-dwarf science (Section~\ref{sec:implications_BD}) and for the selection and validation of ultra-high-redshift galaxies in deep \textit{JWST} fields (Section~\ref{section:implications_galaxy}).

\subsection{Parallax and proper motion}
\label{subsec:parallax&PM}

\begin{figure}
    \centering
    \includegraphics[width=\columnwidth]{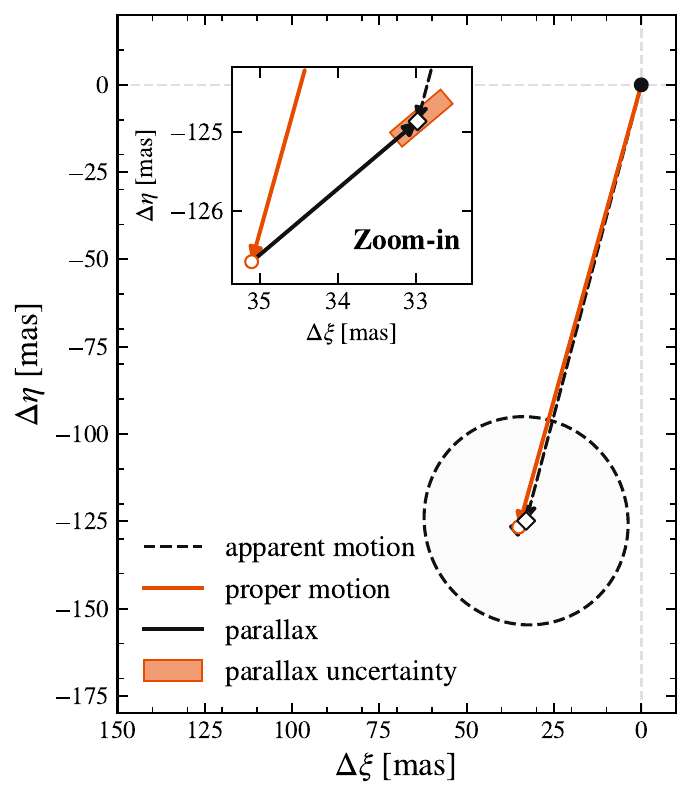}
    \vspace{-0.2in}
    \caption{
    Decomposition of the apparent motion of Capotauro into its proper-motion and parallax components. The dashed black vector shows the measured apparent displacement of Capotauro. The solid black vector shows the expected parallactic displacement computed using the distance inferred from the SED fit, while the orange vector shows the resulting proper-motion component after subtracting this parallax contribution. The inset provides a zoomed view of the endpoint of the decomposition. The dashed circle marks the uncertainty on the measured apparent motion, and the orange rectangle indicates the uncertainty in the parallactic displacement due to the uncertainty on the SED-based distance.
    }
    \label{fig:capotauro_motion_decomposition}
\end{figure}

The SED-derived distance, $d_{\rm SED}=730\pm110$~pc, allows us to isolate the proper motion and parallax components of the total apparent motion, which was not possible from the astrometry alone. We estimated the parallax as $\varpi=1000/d_{\rm SED}$, with distance in pc and parallax in mas. For each of 50,000 astrometric trials in the apparent motion measurement (Sec.~\ref{sec:offset_measurement}), we drew a distance, calculated the corresponding filter-dependent {\it JWST} parallax shift, and removed it from the four measured positions. We then fitted a straight motion in time to the F410M, F444W, F430M, and F460M positions using their full positional covariance.
We obtain an SED-derived parallax of $\varpi_{\rm SED}=1.4^{+0.2}_{-0.2}$~mas, corresponding to an old-to-new displacement of $\Delta\xi_{\rm plx}=-2.1^{+0.3}_{-0.4}$~mas and $\Delta\eta_{\rm plx}=+1.79^{+0.3}_{-0.2}$~mas. After removing this displacement, the eastward and northward proper motions are $\mu_{\xi}=+10.0^{+6.4}_{-4.8}$ and $\mu_{\eta}=-36.2^{+5.1}_{-5.9}$~mas~yr$^{-1}$, giving a total proper motion of $\mu=38.3^{+5.5}_{-5.6}$~mas~yr$^{-1}$ at a position angle of $164.6^{+7.0}_{-9.4}$~deg east of north. Here, as before, $\xi$ and $\eta$ denote the eastward and northward sky coordinates, respectively.

The parallax displacement has a total size of only $2.8^{+0.5}_{-0.4}$~mas, compared with the measured apparent displacement of $131.7^{+19.1}_{-19.6}$~mas. The SED-derived parallax is also fully compatible with the independent astrometry-only result, $\varpi_{\rm ast}=122^{+445}_{-434}$~mas, although that direct measurement is too uncertain to provide a useful parallax constraint. As illustrated in Fig.~\ref{fig:capotauro_motion_decomposition}, the expected parallax is small and points in a different direction from the apparent displacement. Therefore, it cannot account for the observed movement, and  proper motion accounts for almost all of the observed displacement.

\subsection{No evidence for variability}
\label{section:variability}

We also compare the two independent epochs of imaging to test for variability in the brightness of Capotauro. To do this, we first fitted the empirical brown-dwarf templates (Section~\ref{sec:sed}) using only the epoch~2 measurements in F430M, F460M, and F480M, leaving the epoch 1 F444W measurement unused. For each template, we then computed the synthetic F444W flux by convolving the fitted SED with the F444W filter transmission, and averaged over the template uncertainty. This yields a synthetic epoch 2 F444W flux of $S_{\rm F444W}=17.3^{+3.6}_{-3.2}$ nJy, consistent with the directly measured epoch 1 value of $19.4\pm1.8$ nJy. The difference, defined as epoch 2 minus epoch 1, is $-2.1^{+4.0}_{-3.7}$ nJy (or a decrease of $\sim$10\%), corresponding to $\sim0.5\sigma$. Thus, we do not detect any significant change in the brightness of Capotouro in our measurements over the available 3.5-yr baseline.

As an independent test, we fitted a common brown-dwarf spectral template of the same shape separately to each, but allowing the overall flux normalizations to change between epochs, fitting F410M and F444W for epoch~1 and F430M, F460M, and F480M for epoch~2. The resulting flux ratio is $S_{2}/S_{1}=0.87^{+0.19}_{-0.17}$, again formally slightly fainter at the later epoch but fully consistent with unity. Allowing separate epoch normalizations improves the fit by only $\Delta\chi^2=0.68$ for one additional parameter, equivalent to $0.83\sigma$. The new imaging therefore provides no evidence for long-term variability.

Our data cannot, however, rule out a flux change of $\lesssim10\%$ between the two epochs. This is exactly the level at which most variable brown dwarfs are observed to vary in the infrared \citep[0.2\% to 10\%, e.g.,][]{Metchev2015, Vos2022, Biller2024, Liu2024, Cushing2026}, and several substellar objects even show much stronger amplitudes, reaching almost 40\% in one case \citep[e.g.,][]{Radigan2012, Eriksson2019, Bowler2020, Zhou2022, Sutlieff2023, Sutlieff2024, Tan2025}. This could therefore easily explain the possible small decrease in flux between the two imaging epochs. Such variability is typically attributed to inhomogeneous features in the atmospheres of substellar objects that pass in and out of view over their rotation periods, but also evolve dynamically over both rapid and longer-term timescales \citep[e.g.][]{Apai2017, Tan2019, Vos2023, McCarthy2024, McCarthy2025, Plummer2024, Chen2025, Lee2025, OliverosGomez2026}. We stress that our two-epoch sampling constrains only secular, long-term changes: any rotational modulation is averaged over, and effectively randomly phased, within the hours-to-days span of each epoch. A dedicated monitoring campaign would be required to probe any short-term variability of Capotauro at the few-per-cent level or higher.

Another point requiring discussion in this context is the results of \citet{ferrara2026}, who propose the alternative scenario that Capotauro could be a pair-instability supernova \citep[PISN; e.g.,][]{Barkat1967, Rakavy1967, Gilmer2017} at $z\simeq15$ (although they do note that a Y0 brown dwarf solution is also consistent with the observations). While PISNe have been predicted since the 1960s and play an important role in shaping the landscape of black hole formation \citep{Farmer2019, Woosley2021}, only a few candidates have been found to date, including the promising SN~2018ibb \citep[e.g.,][]{Schulze2024}. For Capotauro, a key piece of evidence invoked for this transient origin is an apparent $\sim$23\% brightening between the epoch 1 NIRCam F444W flux of \citet{gandolfi2026} and a synthetic F444W flux extrapolated from the March 2025 NIRSpec spectrum, under the assumption that the NIRSpec spectrum can be flux-calibrated independently of the imaging photometry. As highlighted above, variability at this amplitude is uncommon but not implausible for a strongly variable brown dwarf. However, both of the flux values used to claim this variability warrant caution. First, as shown in Section~\ref{sec:photom_measurement}, the NIRCam fluxes of \citet{gandolfi2026} were likely overestimated by $\simeq50\%$ owing to an erroneously large Kron aperture, so the photometric baseline of the claimed brightening shifts substantially once our revised measurements are adopted.

Second, NIRSpec micro-shutter-assembly (MSA) spectroscopy alone does not deliver robust absolute photometry for a faint point source. The flux entering the $0.20''\times0.46''$ micro-shutters is sensitive to the position of the source within its shutter, on the wavelength-dependent PSF, and to diffraction by the shutter itself, so the recorded flux is suppressed by slit losses that are difficult to model from first principles \citep{Jakobsen2022, Ferruit2022}. For this reason, the absolute flux scale of MSA spectra is routinely anchored to imaging: synthetic photometry computed from the spectrum is matched to the measured broad-band fluxes, either with a constant factor or with a low-order wavelength-dependent rescaling \citep[e.g.,][]{ArrabalHaro2023, Reddy2023, Cameron2023, 2026MNRAS.547ag449S, Leung2026}. Independent estimates of this scaling can differ by 10 to 40\% for individual sources \citep{ArrabalHaro2023}, and sometimes up to $100\%$ \citep{2024ApJ...976..193R}. A NIRSpec-based flux that is not tied to contemporaneous imaging therefore cannot establish such a flux change; conversely, if the spectrum were rescaled to match earlier photometry, the comparison would carry no independent information on variability.

The absence of variability between the two NIRCam imaging epochs may be inconsistent with the $z\simeq15$ PISN interpretation; our 3.5 yr baseline corresponds to $\sim$80 rest-frame days at $z\simeq15$, an interval over which a luminous transient would generically be expected to evolve in brightness, even allowing for the strong time dilation. For example, the candidate PISN SN~2018bb shows variability over this (rest-frame) time scale \citep{Schulze2024}. Some models of PISNe do predict constant luminosities for more than 100 (rest-frame) days after peak \citep{Kasen2011, Kozyreva2014, Gilmer2017}. These typically concern the most extreme cases from the most massive (and thus rarer) progenitors with initial masses higher than 130\Msun. Taken together, the revised photometry, the absence of detectable variability, and the significant proper motion consistently favour the brown-dwarf interpretation of Capotauro.

\subsection{Implications for future work on brown dwarfs}
\label{sec:implications_BD}

The measured proper motion and SED establish Capotauro as a Y~dwarf with a photometric spectral type of Y$1.0\pm0.5$, corresponding approximately to $\Teff\simeq350$\,K. If Capotauro is a single object with an intrinsic luminosity similar to the best-fitting templates, its F444W flux implies a distance of $730\pm110$\,pc. This implies a galactic coordinate of $Z=630\pm100$\,pc, placing it within about two thin-disc scale heights, or one thick-disc scale height \citep{Vieira2023}. We therefore would not expect it to necessarily belong to a population distinct from solar neighborhood field objects. Its kinematics may provide additional insight, and when combined, the distance and proper motion yield $v_{\rm tan}=132\pm28\,{\rm km\,s^{-1}}$. This relatively large tangential velocity may indicate an old, thick-disc object, but the uncertainty is quite large, so we must consider all possibilities. 

We generated a Besan\c{c}on galaxy model \citep{Robin2003,Robin2017} at Capotauro's coordinates and examined the properties of simulated stars along the line-of-sight up to distances of 2\,kpc. We neglect the detailed stellar properties of the simulated stars and use them simply as kinematic tracers, considering only their ages and velocities. We used rejection sampling to trim the simulation output to have distances distributed according to our estimate of $730\pm110$\,pc. The simulated stars are almost exactly evenly divided between the thin disc (51\%, ages 1--9\,Gyr) and thick disc (49\%, ages 10--12\,Gyr). Halo stars are negligible, making up $<$0.5\% of the simulation. To incorporate our measurement uncertainties, we added 30\,km\,s$^{-1}$ Gaussian noise to the simulated tangential velocities. Even after this, such high velocities as we observe are uncommon both in the thin- and thick-disc samples, with probabilities of 0.4\% and 8\%, respectively. This suggests that Capotauro may be an older brown dwarf belonging to the thick disk population. Alternatively, as we noted previously, Capotauro could be somewhat colder and fainter than covered by the available templates, which would mean its distance, and thus inferred velocity, would likely be smaller, thereby increasing the chance of it being a field-like thin-disc object.

Capotauro itself holds potential interest for substellar science, given how few objects are still known at the cold temperatures with which Capotauro's photometry is consistent. Y~dwarfs' $3$--$12\,\micron$ spectra are sensitive to molecules such as CH$_4$, CO$_2$, H$_2$O, NH$_3$, and PH$_3$, which can probe physical processes such as disequilibrium chemistry, vertical mixing, and cloud rainout, as well as the fundamental parameters of metallicity, surface gravity, and C/O ratio \citep[e.g.,][]{Morley2014,Beiler2024,Mukherjee2024,Leggett2025,Leggett2026}. The $4$--$5\,\micron$ spectral region is particularly sensitive to gravity, metallicity, and disequilibrium chemistry \citep{Beiler2024,Mukherjee2024,Leggett2025}, which presents both a challenge and an opportunity for an object like Capotauro that has so far been detected only in this wavelength range. Its fundamental properties may be more difficult to isolate, but its SED also probes some of the most poorly understood substellar atmospheric physics. Ultimately, Capotauro will be as valuable to substellar science as the data it can provide, which, given its faintness, presents another challenge. Without deeper data, and additional imaging epochs to constrain its distance via trigonometric parallax, Capotauro will primarily be of interest as a statistical probe of the poorly known population of very distant brown dwarfs.

At the population level, finding a brown dwarf as faint, cold, and distant as Capotauro is a significant milestone. It demonstrates that deep, {\it JWST} pencil-beam surveys can potentially constrain the scale height, kinematics and low-mass end of the Galactic substellar mass function well into the regime of Y~dwarfs. The serendipitous example of Capotauro suggests that broad-band imaging over a wide range of wavelengths combined with medium-band imaging over the 4--5\,\micron\ range is ideal for this purpose.

\subsection{Implications for future work on high-redshift galaxies}
\label{section:implications_galaxy}

The confusion between Capotauro and a putative $z\simeq32$ galaxy is a new manifestation of a long-standing problem in Lyman-break galaxy selection. In ground-based and \textit{HST}-era searches at $z\simeq5$--7, cool M, L and especially T~dwarfs could reproduce the red-optical-to-near-infrared colours of compact high-redshift galaxies (see \citealt{Dunlop2013} for a review). \citet{Bowler2014,Bowler2015,Bowler2020} explicitly fitted cool-dwarf templates, together with deep optical and \textit{Spitzer}/IRAC data, to reject stellar contaminants from bright high-redshift galaxy samples based primarily on ground-based seeing-limited data. Our result provides a direct empirical demonstration of this general concern and also shows how it has shifted into a new regime: the relevant contaminants are now colder Y~dwarfs, whose spectra were absent from many earlier stellar libraries and whose strongest observable fluxes are confined to the longest-wavelength NIRCam filters.

A Y-dwarf SED is expected to create interlopers clearly separated from the current galaxy redshift frontier at $z\sim15$. The broad H$_2$O, CH$_4$ and NH$_3$ absorption \citep{Beiler2024, Leggett2025} in a $\Teff\simeq300$--400\,K Y~dwarf can suppress essentially all measurable flux through F356W, before the SED rises sharply near $4$--$5\,\micron$. If this atmospheric rise is interpreted as a Lyman break, its position at $\simeq3.8$--$4.0\,\micron$ maps naturally to $z_{\rm phot}\simeq30$--32. Thus the coldest stellar contaminants may produce a separate accumulation of apparent F356W-dropout candidates near $z_{\rm phot}\simeq30$. Capotauro, the two moving Y~dwarfs in the Bullet-cluster field \citep{bradac2026}, and the brown dwarfs to be presented by F.~Liu et al. (in prep.) provide direct examples of this tendency.

In the context of the continuing lack of secure spectroscopic confirmations at $z>15$, and emerging evidence for a rapid decline in the galaxy UV luminosity and star-formation-rate densities beyond $z\simeq12$ \citep{Weibel2026,mcleod2026}, these results warrant particular caution when interpreting unresolved $z_{\rm phot}\gtrsim30$ candidates supported by only a few long-wavelength detections. As demonstrated by the case of Capotauro, multi-epoch imaging can provide clean discrimination via relative astrometry, and repeated imaging can also reject transient contamination, as illustrated by the disappearance of the proposed $z\simeq14$ BEACON source in independent epochs \citep{donnan2026,kreilgaard2026,toshikage2026}. Therefore, in the {\it JWST} era, to efficiently and robustly select ultra-high-redshift galaxies, we are now in need of sufficiently cold Y-dwarf templates, deep medium-band photometry, and multi-epoch astrometry.

\section{Conclusions}
\label{section:conclusion}

We have performed a multi-epoch astrometric and photometric study of Capotauro, the compact F356W-dropout source in the CEERS field \citep{Finkelstein2022, Finkelstein2023, Bagley2023} proposed as a $z_{\rm phot}\simeq32$ galaxy candidate by \citet{gandolfi2026}, for which a very cool Y-type brown dwarf has always been the principle competing interpretation. New NIRCam medium-band imaging from the MINERVA \citep{muzzin2025} and SPAM programs, obtained $\simeq3.5$ years after the discovery data, provides both the time baseline needed to measure apparent motion and new photometry sampling the SED, allowing the two hypotheses to be tested directly. Our main conclusions are as follows:

\begin{itemize}
    \item We detect an apparent displacement of $131.7^{+19.1}_{-19.6}$ mas between the two epochs at $6.2\sigma$ significance, corresponding to $(\mu_{\xi}, \mu_\eta) = (+9.4^{+6.4}_{-4.8}, 35.6^{+5.1}_{-5.9})$ mas yr$^{-1}$, or $37.6^{+5.5}_{-5.6}$ mas yr$^{-1}$ in total. This rules out any extragalactic interpretation, i.e., both the $z\simeq32$ galaxy \citep{gandolfi2026} and the $z\simeq15$ pair-instability supernova \citep[PISN,][]{ferrara2026} scenarios. 

    \item Our measured photometry is best matched by an empirical Y$1.0\pm0.5$ template with $\Teff\simeq350$\,K and $\log(\Lbol/\Lsun)=-6.68\pm0.11$, at a template-scaled distance of $730\pm110$\,pc. This is consistent with the results found by \citet{hainline2026b} for Capotauro by fitting the December 2022 NIRCam and March 2024 MIRI photometry from \citet{gandolfi2026} despite the likely overestimation of the \citet{gandolfi2026} NIRCam photometry. However, as the empirical library is restricted to spectral types of Y1 and earlier, these values are effectively limits and an even colder, closer object with a later spectral type cannot be excluded. The parallax-corrected proper motion implies $v_{\rm tan}= 132\pm28$\,km~s$^{-1}$, consistent with both thin- and thick-disk kinematics, making Capotauro a valuable probe of the low-mass end of the substellar mass function and of an atmosphere comparable in temperature to those of the coldest directly imaged exoplanets.

    \item The Capotauro-like Y-type brown dwarfs can masquerade as a population of Lyman break galaxies at $z_{\rm phot}\simeq30$. Therefore, robust selection of ultra-high-redshift candidates should explicitly incorporate cold brown dwarf templates together with galaxy templates, and examine relative astrometry when multi-epoch images exist.
\end{itemize}

The present data leave critical questions about Capotauro open. A third astrometric epoch timed for probing its parallax will properly constrain its geometric distance. In an ideal world, NIRSpec and/or MIRI spectroscopy would facilitate the expansion of current empirical BD templates into cooler types through Capotauro. However, its faintness makes such observations inherently impractical. Instead, the discovery of Capotauro as a cool Y dwarf indicates that new and wider \textit{JWST} broad-band imaging over a wide wavelength range in tandem with medium-band imaging over the key 4--5\,\micron\ range can yield the discovery of other similar but brighter objects. These objects will be prime opportunities for deep NIRSpec and/or MIRI follow up observations, which will effectively probe currently poorly-understood substellar physical processes and chemistry.

\section*{acknowledgements}
DJM and JSD acknowledge the support of the Royal Society through the award of a Royal Society University Research Professorship to JSD. BJS acknowledges funding by the UK Science and Technology Facilities Council (STFC) grant nos. ST/V000594/1 and UKRI1196. EL acknowledges support through a Veni grant (VI.Veni.232.205) from the Netherlands Organization for Scientific Research (NWO), and the Research Foundation – Flanders (FWO) under the Odysseus Program, Type II (G0AT525N).

This work made use of the University of Edinburgh's Institute for Astronomy research computing cluster, {\sl cuillin}, which is partially funded by the UK STFC. We would like to thank Eric Tittley for his work managing and maintaining {\sl cuillin}. This work is based in part on observations made with the NASA/ESA/CSA James Webb Space Telescope. The data were obtained from the Mikulski Archive for Space Telescopes (MAST) at the Space Telescope Science Institute (STScI), which is operated by the Association of Universities for Research in Astronomy, Inc., under NASA contract NAS 5-03127 for {\it JWST}. These observations are associated with program nos. ERS~1345, GO~2234, GO~3794, GO~7814, \& GO~8559. The authors acknowledge the ERS~1345, GO~7814 and GO~8559 teams for developing their observing programs with a zero-exclusive-access period. This work has made use of data from the European Space Agency (ESA) mission {\it Gaia} (\url{https://www.cosmos.esa.int/gaia}), processed by the {\it Gaia} Data Processing and Analysis Consortium (DPAC, \url{https://www.cosmos.esa.int/web/gaia/dpac/consortium}) \citep[][]{2016A&A...595A...1G, 2023A&A...674A...1G}. Funding for the DPAC has been provided by national institutions, in particular the institutions participating in the {\it Gaia} Multilateral Agreement. This research has made use of the Spanish Virtual Observatory (SVO) Filter Profile Service ``Carlos Rodrigo" \url{https://svo2.cab.inta-csic.es/theory/fps/}, funded by MCIN/AEI/10.13039/501100011033/ through grant PID2023-146210NB-I00 \citep{Rodrigo2012, Rodrigo2020}. This research has made use of the SIMBAD database, operated at CDS, Strasbourg, France \citep{2000A&AS..143....9W}. This research made use of SAOImageDS9, a tool for data visualization supported by the Chandra X-ray Science Center (CXC) and the High Energy Astrophysics Science Archive Center (HEASARC) with support from the {\it JWST} Mission office at the Space Telescope Science Institute for 3D visualization \citep{2003ASPC..295..489J}. This research has made use of the Astrophysics Data System, funded by NASA under Cooperative Agreement 80NSSC21M00561. This work makes use of the Python programming language\footnote{Python Software Foundation; \url{https://www.python.org/}}, in particular packages including \texttt{Astropy} \citep{astropy, Astropy2018, Astropy2022}, \texttt{Matplotlib} \citep{matplotlib}, \texttt{Numpy} \citep{numpy}, pandas \citep{mckinney-proc-scipy-2010, jeff_reback_2022_6408044}, \texttt{Photutils} \citep{photutils}, \texttt{Scipy} \citep{scipy}, and \texttt{STPSF} \citep{Perrin2012, Perrin2014}.

\section*{Data Availability}
The data underlying this article will be shared on reasonable request to the corresponding author.

\bibliographystyle{mnras}
\bibliography{Capotauro_Liu_2026.bib}

\bsp	
\label{lastpage}

\end{document}